\documentclass[aps,prl,twocolumn,floatfix,superscriptaddress,showpacs,longbibliography]{revtex4-1}
\usepackage{graphicx}
\usepackage{amsmath}
\usepackage{amssymb}
\usepackage{amsfonts}
\usepackage{bm}        
\usepackage{dsfont}
\usepackage[mathscr]{eucal}
\usepackage[colorlinks, linkcolor=OrangeRed,citecolor=RoyalBlue,urlcolor=NavyBlue]{hyperref}
\usepackage[normalem]{ulem}
\usepackage[all]{hypcap}
\usepackage[dvipsnames,usenames,table]{xcolor}

\newcommand{\vk}{\mathbf{k}}

\newcommand{\eq}[1]{\begin{align}#1\end{align}}

\newcommand{\nn}{\nonumber}
\newcommand{\pr}{\mathbf{r}}
\newcommand{\msr}[1]{\mathscr{#1}}
\newcommand{\mcl}[1]{\mathcal{#1}}
\newcommand{\mrm}[1]{\mathrm{#1}}

\begin{document}
\title{Chern numbers and chiral anomalies in Weyl butterflies}

\author{Sthitadhi Roy}
\affiliation{Max-Planck-Institut f{\"u}r Physik komplexer Systeme, N{\"o}thnitzer Stra{\ss}e 38, 01187 Dresden, Germany}
\author{Michael Kolodrubetz}
\author{Joel E. Moore}
\affiliation{Materials Sciences Division, Lawrence Berkeley National Laboratory, Berkeley, CA 94720, USA}
\affiliation{Department of Physics, University of California, Berkeley, CA 94720, USA}
\author{Adolfo G. Grushin}
\affiliation{Department of Physics, University of California, Berkeley, CA 94720, USA}

\begin{abstract}
The Hofstadter butterfly of lattice electrons in a strong magnetic field is a cornerstone of condensed matter physics,
exploring the competition between periodicities imposed by the lattice and the field. 
In this work we introduce and characterize the Weyl butterfly, which emerges when a
large magnetic field is applied to a three-dimensional Weyl semimetal. 
Using an experimentally motivated lattice model for cold atomic systems, we solve this problem numerically.
We find that Weyl nodes reemerge at commensurate fluxes and propose using wavepackets dynamics to reveal their chirality and location. 
Moreover, we show that the chiral anomaly -- a hallmark of the topological Weyl semimetal -- 
does not remain proportional to magnetic field at large fields, but rather inherits a fractal structure of linear regimes as a function of external field.
The slope of each linear regime is determined by the difference of two Chern numbers in gaps of the Weyl butterfly 
and can be measured experimentally in time-of-flight. 
\end{abstract}
\maketitle


\paragraph{Introduction -- }
The search for novel phenomena in condensed matter is often spurred by
competing physical effects acting on comparable length scales. 
A central example is the response of fermions on a lattice to an external magnetic field when 
the magnetic length $l_{B}$ becomes comparable to the lattice spacing $a$.
The energy spectrum displays a fractal structure known as the Hofstadter butterfly~\cite{H76}, that is
invariant upon changing applied field by one magnetic flux quantum $\Phi_{0}=h/e$
per unit cell. 
Furthermore, the gaps that separate sub-bands are characterized by a two-dimensional topological Chern number.

Topological invariants generally yield non-trivial condensed matter phenomena, as in
topological insulators~\cite{HK10,QZ11} and more recently Weyl and Dirac semimetals~\cite{Hosur:2013eb,Turner:2013tf,Vafek:2014hl}. 
These semimetals host pairs of protected band touchings (nodes) that disperse linearly with momentum.  
Each pair is composed of a left and right chirality node, a quantum number resembling the valley degree of freedom in graphene.
%
\begin{figure}
 \includegraphics[width=\columnwidth]{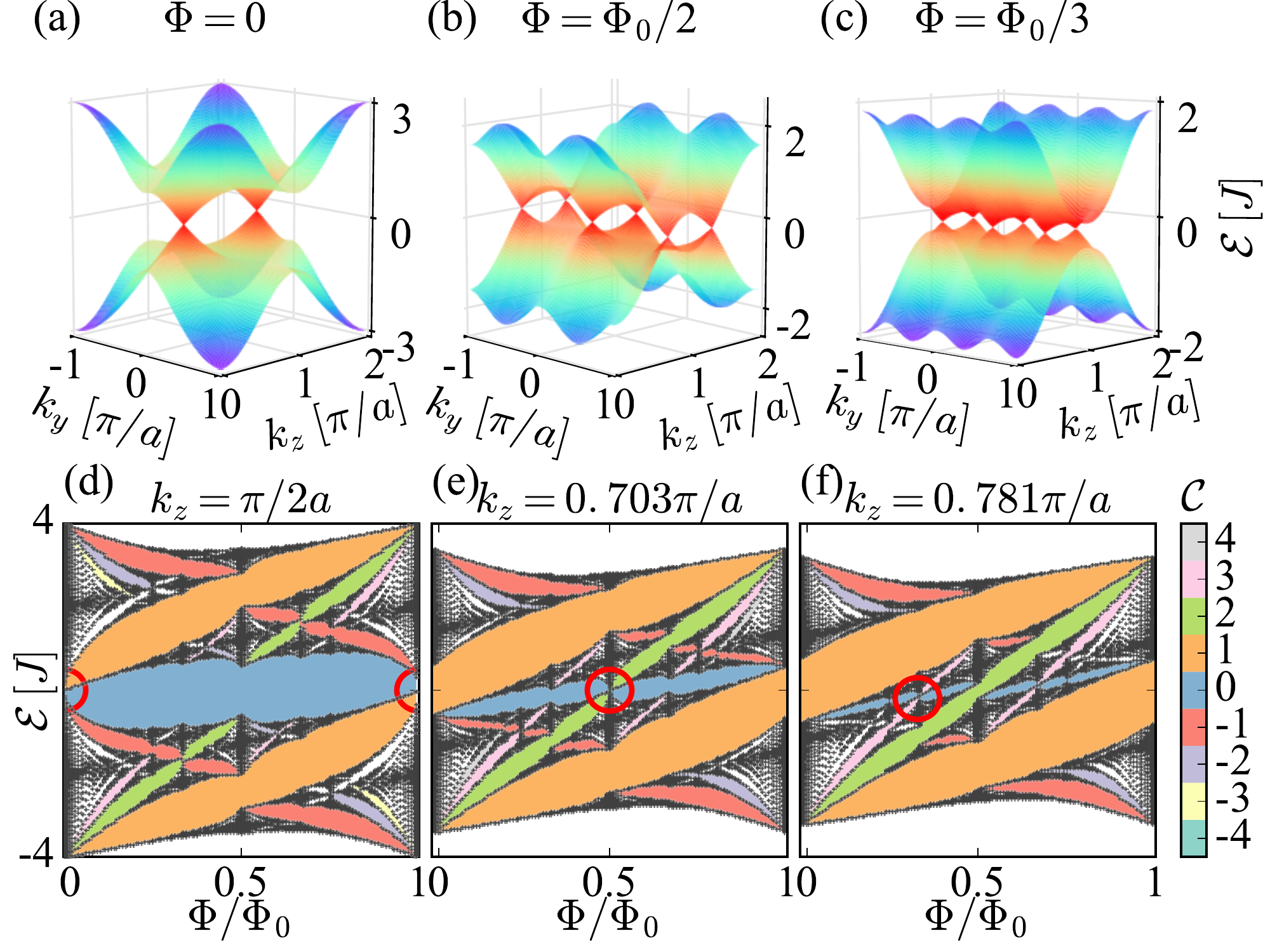}
 \caption{Energy spectra of Weyl fermions in a magnetic field, as obtained from Eq.~\eqref{eq:ham_hof}. {\bf(a)-(c)} New Weyl nodes emerge 
at commensurate fluxes per plaquette with an additional $q$-fold degeneracy (see text). 
{\bf(d)-(f)} Hofstadter-like spectrum for the $k_z$ values where the Weyl nodes appear for the 
corresponding fluxes shown in (a)-(c). The isolated zeroth Landau level
occurring around the Weyl nodes is highlighted by the red circles. 
The colors in the butterflies denote the Chern number $\mathcal{C}$ in the gap.
}
 \label{fig:weyls_butterflies}
\end{figure}
Although the sum of right- and left-handed fermions is conserved, non-orthogonal magnetic ($\mathbf{B}$) and electric ($\mathbf{E}$) fields 
pump one chirality to the other at a rate $\propto \mathbf{E}\cdot \mathbf{B}$, so that their difference is not conserved.
This phenomenon, known as the chiral anomaly~\cite{bertlmann2000anomalies,Nielsen1983}, distinguishes conventional metals from topological ones;
while the former display positive longitudinal magnetoresistance, the chiral anomaly manifests as negative magnetoresistance in the latter~\cite{Son:2013jz,Burkov:2015wt},
consistent with recent  measurements~\cite{Xu2013,KimiKim2013,Feng2014,Borisenko2014,Li2014,Liu2014a,Neupane2014,Yi2014,Liu2014b,He2014,Huang2015b,ZhangXu2015,XKL2015,YangLi2015,SAW2015,Du2015,ZGC2015,LiHe2015,Weng2015,Huang2015,SIN15,Lv2015,XAB15,LvXu2015,YLS15,SJB15,Xu2015,Liang2015}.

\begin{figure}
\includegraphics[width=\columnwidth]{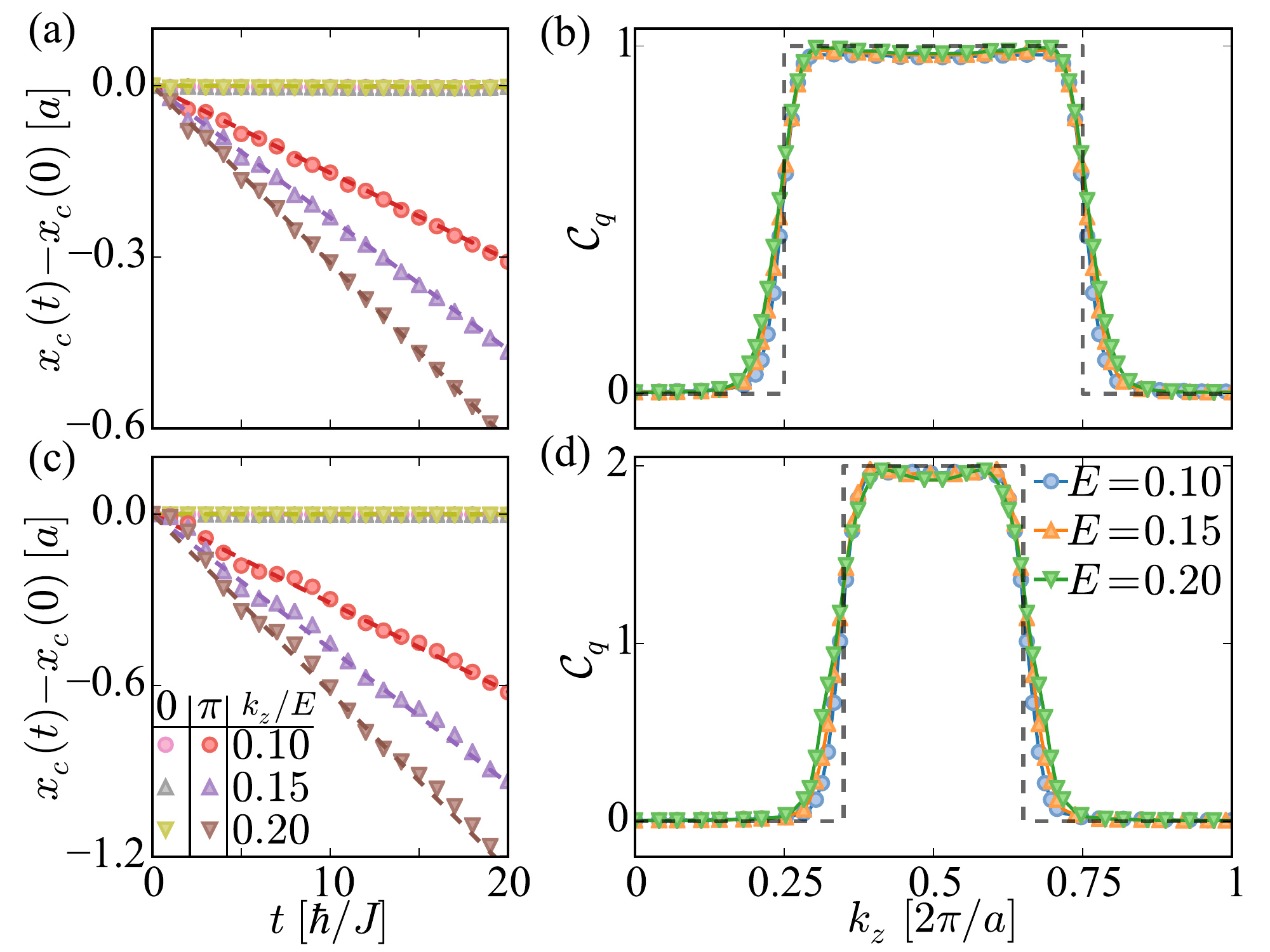}
\caption{ Hall drift of the center-of-mass of a wavepacket for different values 
of the electric field $E$ at two representative  $k_z$ values, 
$0$ and $\pi$, along with the linear fits used to calculate $\mcl{C}_q$ for $q=1$ 
{\bf(a)} and $q=2$ {\bf(c)}. {\bf(b)} and {\bf(d)} show the topological transition of
$\mcl{C}_q$ at the positions of the Weyl nodes, as obtained from Eq.~\eqref{eq:drift}. 
For the simulations $L=128$ and $L_s=48$.
}
 \label{fig:chern}
\end{figure}

The response of Weyl fermions to external electromagnetic fields is well understood both in the linear response regime~\cite{volovik_book,Burkov:2011de,Zyuzin:2012ca,Aji:2012gs,Grushin2012,Liu:2013kv,Goswami:2013jp,Landsteiner:2014fw} and the Landau level limit~\cite{Nielsen1983,Goswami:2013jp,KGM15,BQ16,OK15}.
%
%
In this work, we uncover a richer structure that emerges for Weyl fermions on a lattice in the Hofstadter regime ($l_B\sim a$), determining the fate of the chiral anomaly in this limit. 
%
The non-renormalization theorem of the chiral anomaly beyond one-loop~\cite{bertlmann2000anomalies}
breaks down, 
and the $\mathbf{B}$-dependence of the anomaly is periodic in units of $\Phi_{0}$ per unit cell.
%
%
%
%
We find that the chiral anomaly tracked through the rate of chiral charge pumping shows a fractal of linear regimes proportional to $\mathbf{B}$ with quantized integer slopes, intimately connected to a fractal set of emergent Weyl nodes at commensurate fluxes.
%
%
The integer slopes are given by the Chern numbers of the Weyl butterfly -- a three-dimensional fractal which describes the spectrum of a Weyl semimetal under large magnetic fluxes.
%
The physics resulting from the third dimension is not a mere generalization of the two-dimensional Hofstadter case.
%
%
Crucially, we show that there is an analytical connection between the evolution of the fractal spectrum along the third momentum direction, its Chern numbers, 
and the fractal nature of the chiral anomaly.
The work motivated by both fundamental interest~\cite{KHW92,H92,KAK01,KAO02,GK02,KA03,KA04,BDG04} and plausible experimental realization in cold atomic systems~\cite{DKLKSB2015,ZZW2015,HZL2015}.
Using a model tightly connected to the experimental proposal in Ref.~\cite{DKLKSB2015} 
we show how both the Chern numbers and the chiral anomaly may be directly measured with cold atoms.
%
%
%
%
\paragraph{Weyl semimetal in magnetic fields --}
To describe a Weyl semimetal, we employ a two-band Hamiltonian of spinless fermions on a cubic lattice $\msr{H}_\vk = \mathbf{d}_\vk\cdot\bm{\sigma}$, with
\eq{
\mathbf{d}_\vk=-\{&J_2 \sin k_x,J_2\sin  k_y,-M+J_1\sum_{i=x,y,z}\cos k_i\}.
\label{eq:ham_mom}
}
This model breaks time-reversal symmetry and is
motivated by that in Ref.~\cite{DKLKSB2015}, which is composed of the two time reversal partners of Eq.~\eqref{eq:ham_mom} separated in momentum space.
It has a pair of linearly dispersing Weyl cones at $\vk=\{0,0,\pm\cos^{-1}(M/J_1-2)\}$ for $1<\vert M/J_1\vert<3$.
We use $J_2=J_1=J$ and $M/J=2$ for the remainder of this paper.
Fig.~\ref{fig:weyls_butterflies}(a) shows the band structure in the $(k_y,k_z)$ plane.
This model is constructed from Chern insulators in the $(k_x,k_y)$ plane 
with a $k_z$ dependent gap such that the Chern number $\mcl{C}_{k_z}$ changes whenever a Weyl node is crossed, as shown
by the dashed line in Fig.~\ref{fig:chern}b.
We now consider applying magnetic field $\mathbf{B}\parallel \hat{z}$ with flux
$\Phi = \Phi_0 p/q$ per plaquette, equivalent to an Aharonov-Bohm phase of $\phi=2\pi p/q$ upon tunneling around a plaquette.
In the Landau gauge, $\mathbf{A} = \Phi x\hat{y}$, the Hamiltonian becomes 
\eq{
\nn&\msr{H}_{\msr{WH}} = \begin{pmatrix}
              \msr{M}_1&\msr{R}/2&0&\cdots&\msr{S}/2\\
              \msr{R}^\dagger/2&\msr{M}_2&\msr{R}/2&\cdots&0\\
              \vdots&\ddots&\ddots&\ddots&\vdots\\
              \msr{S}^\dagger/2&0&\cdots&\msr{R}^\dagger/2&\msr{M}_q\\
              \end{pmatrix},\\
\nn&\msr{M}_n(k_y,k_z) =~[M-J(\cos(k_y+\phi n) + \cos k_z)]\sigma^z -\\
\nn&~~~~~~~~~~~~~~~~~~~~~~~~~~~~~J\sin(k_y+\phi n)\sigma^y,\\
\msr{R}&=-J(\sigma^z-i\sigma^x);~ \msr{S}(k_x)=-J(\sigma^z+i\sigma^x)e^{-ik_x}.
\label{eq:ham_hof}
}
Each $k_z$ exhibits a Hofstadter-like spectrum~\cite{AC14}, (see Fig.~\ref{fig:weyls_butterflies}) 
forming the energy spectrum that we refer to as the Weyl butterfly.

One unexpected aspect of the Weyl butterfly is that, for commensurate fluxes,  
new pairs of Weyl nodes emerge with $q$-fold degeneracy unlike the two-dimensional Hofstadter butterfly which has Dirac nodes only for even $q$.~\footnote{By inspection of the spectrum it seems likely that given a commensurate value of $\Phi/\Phi_0=p/q$
there is a particular value of $k_z$ where $q$ Weyl nodes appear somewhere in the spectrum (not necessarily at E=0). Unfortunately we have no rigorous proof and thus 
leave it as a conjecture (see supplementary information).}~\cite{WZ1989,LFTB2015}
In Fig.~\ref{fig:weyls_butterflies}(d)-(f) we highlight some of the emergent Weyl nodes that cross near $\mathcal{E}=0$ for particular values of $k_z$.
The emergence of such new Weyl nodes is related to the fractal structure of the butterfly, while the
$q$-fold degeneracy comes from noting that the shift $k_y{\rightarrow}k_y{+}2\pi p/q$ 
in Eq.~\eqref{eq:ham_hof} amounts to changing $\mathbf{A}$ in steps of $\Phi$, which has no effect on the spectrum.
Since $q$ such translations traverse the BZ, there should be $q$ copies of the spectrum.
\footnote{More generally, the degeneracy will be some multiple $nq$ of $q$ for integer, since we know
that if $n$ Weyl pairs exist, then we can create $n q$ copies by shifting $k_y$. For all the cases
considered in this paper, $n=1$.}
Perturbing the flux around one of these emergent Weyl nodes splits it into Landau levels dispersing along $k_z$, including a chiral zeroth Landau level~\cite{Nielsen1983}.
As the flux is further increased, the Landau levels split and merge with those of the upcoming Weyl node,
a feature which we explore in more detail later.
%
%
%
\paragraph{Chern numbers via wavepacket dynamics -- }
%
%
%
A non-trivial topological invariant that characterizes the emergent Weyl nodes at rational flux 
is the Chern number in each momentum plane.
An experimentally feasible probe of this invariant in cold-atomic systems is the semi-classical motion of wavepackets~\cite{DG13,LWZ16}, 
which has successfully been used in both the two-dimensional Hofstadter~\cite{ALS14} and Haldane~\cite{JMD14} models.
The principle is that, under an external force, wavepackets Hall drift transverse to the direction of the force with amplitude proportional to the Chern number.
%
%

%
Here we explore the use of such Hall-like response to characterize the Weyl butterfly.
First, we consider preparing a wavepacket sharply peaked around a finite momentum $k_z$ along the axis of Weyl node separation. 
Such a wavepacket could be achieved experimentally by initially decreasing the lattice depth along the $z$-direction to create a 
sharp $k_z$ peak then taking it to the desired $k_z$ via a ramped magnetic field~\cite{GMEHB2002} 
or optical gradient~\cite{ALS14}, lattice acceleration~\cite{DPRCS1996}, or Bragg pulse.\cite{EGKPLPS2009}
We then consider a wavepacket initially confined within a $L_s\times L_s$ sub-region of an $L\times L$ lattice ($L_s<L$) in the $(x,y)$ plane~\cite{DG13}.
At $t=0$, the $xy$-confinement is removed to give approximately uniformly-filled bands in $(k_x,k_y)$~\footnote{A
uniform population of the band is expected if the hierarchy of energy scales $\Delta\gg k_{B}T \gg W$
is satisfied, where $W$ is the bandwidth $\Delta$ is the band gap and $k_{B}T$ is the temperature
energy scale. This criterion is satisfied for high flatness ratios, which have been demonstrated in experiment~\cite{ALS14}}
and a constant force $\mathbf{F} = EJ/a~\hat{y}$ is applied~\cite{ALS14,GMEHB2002,DPRCS1996}
The center-of-mass motion in the $n^\mrm{th}$-band is governed by the semiclassical equations of motion \cite{Xiao2010}
\eq{
\dot\pr_c =\bm{\nabla}_\vk\mathcal{E}_{n,\vk} - \dot{\vk}\times\bm{\Omega}_{n,\vk};~~~~ \dot{\vk}=\mathbf{F};
\label{eq:semiclassic}
}
where $\mcl{E}_{n,\vk}$ and $\bm{\Omega}_{n,\vk}$ are respectively the energy dispersion and Berry curvature of the band.
The net drift of the many-fermion wavepacket can be obtained by integrating Eq.~\eqref{eq:semiclassic} over time and summing over the responses for all the filled bands.
For $m$ uniformly filled bands, the drift is~\cite{DG13}
\eq{
\pr_c(t) - \pr_c(0) =  -\frac{Et}{2\pi}\sum_{n=1}^m\msr{C}_n~\hat{x}  \equiv -\frac{Et}{2\pi} \mcl{C}_m~\hat{x},
\label{eq:drift}
}
where 
$\msr{C}_n = (1/2\pi)\int dk_x dk_y~ \Omega_{n,\vk}^z$ is the Chern number of the $n$-th band.

For flux $\Phi$, we can use this technique to measure the sum of the Chern numbers of the $q$ lowest bands, $\mcl{C}_q$, for
emergent Weyl nodes which connect the $q^{\mrm{th}}$ and $(q+1)^{\mrm{th}}$ bands.
The Hall drift given by Eq.~\eqref{eq:drift}, and its corresponding $C_q$  are shown in Fig.~\ref{fig:chern} for two different fluxes as a function of $k_z$.
As $k_z$ crosses a Weyl node, the $q^{\mrm{th}}$ and $(q+1)^{\mrm{th}}$ bands undergo a topological phase transition where
the sign of the gap of the Chern insulator flips.
Consequently, since there are $q$ such Weyl points, the Chern number changes by $\pm q$ with sign determined by the chirality of the Weyl nodes.
Thus from the wavepacket dynamics as a function of $k_z$, one can extract the location, chiralities, and multiplicities of the Weyl points.
In experiments, it is often easier to prepare a finite-width distribution of the occupations $W_{k_z}$
than a sharply-peaked $k_z$.
Controlling the width of this distribution
through external trapping or temperature also allows to infer information about the spectrum.
In this case, the Hall drift with the Fermi level in the $q^\mrm{th}$ gap yields a non-quantized effective Chern number, $\mcl{C}_{q,\mrm{eff}}=\sum_{k_z}\mcl{C}_{q,k_z}W_{k_z}$, which is the average of $\mcl{C}_{q,k_z}$ weighted by $W_{k_z}$.
For instance, if we create Gaussian distributions centered at $k_z=0$ with width $\sigma$
for the particular case of two Weyl nodes at $\pm{K_{0}/2}$, 
we find that the dependence of the Hall drift on $\sigma$ saturates to 
$\mcl{C}_{q,\mathrm{eff}}=\mcl{C}_{q,k_{z}=\pi/a}-(\mcl{C}_{q,k_{z}=\pi/a}-\mcl{C}_{q,k_{z}=0})K_{0}/2\pi$
in the $\sigma\to\infty$ limit and to $\mcl{C}_{q,\mathrm{eff}}=\mcl{C}_{q,k_z=0}$ for $\sigma \to 0$.
Varying $\sigma$ interpolates between these limits; a simple fit can then extract the Chern number profile.
A more in-depth analysis can be found in the Supplementary material~\cite{Sup}.
%
%
\paragraph{Chiral anomalies in the Weyl butterfly -- }
%
%
%
\begin{figure}
 \includegraphics[width=\columnwidth]{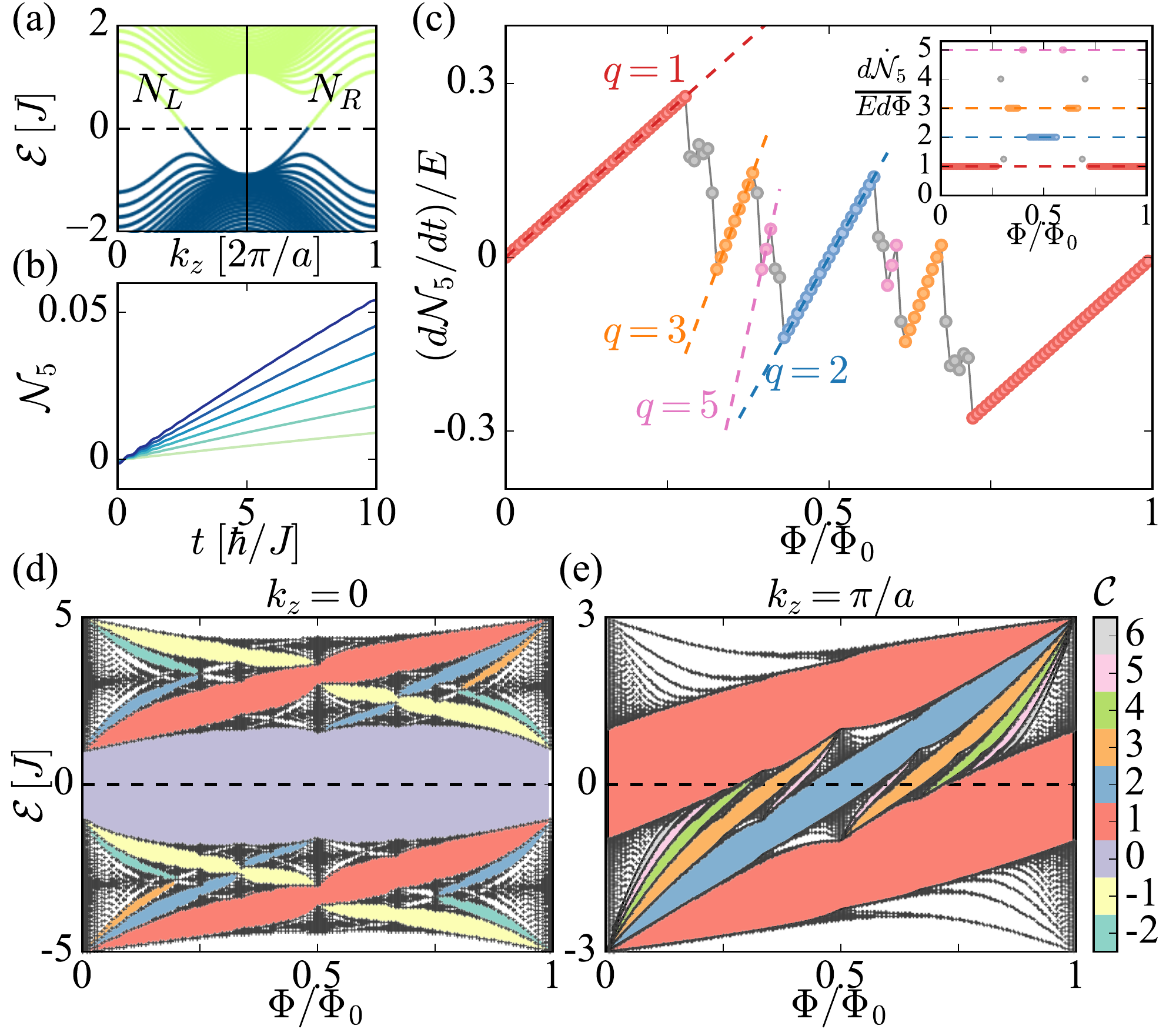}
 \caption{Chiral anomaly of the Weyl butterfly. {\bf(a)} The chiral charge counts the number difference of 
left and right movers $\mcl{N}_5 = (N_L-N_R)/L^2$. {\bf(b)} $\mcl{N}_5$ increases linearly with time when both $E$ and $B$ 
are applied as shown for $E=0.1$ and fluxes $\Phi=n\Phi_0/L$, with $n$ going from 1 to 6 (lighter to darker).  
{\bf(c)}The rate of chiral charge production grows linearly with the flux with a slope $q$ in the
vicinity of commensurate fluxes, for which the model hosts $q$ pairs of Weyl nodes. 
Linear fits are shown for the data around $\Phi/\Phi_0=1,1/2,1/3,2/5$ which correspond to $q=1,2,3,5$ respectively.
Inset: Quantized plateaus corresponding to each linear rate as a function of $\Phi$.
The simulations are performed on a cubic lattice with linear dimension $L{=}144$.
{\bf(d,e)} The magnitude of each plateau can be extracted 
as described by \eqref{eq:Cdiff}.
The slopes in {\bf{(c)}} are color coded to the Chern numbers in {\bf{(d-e)}}
}
 \label{fig:ca_slopes}
\end{figure}
%
We now turn to the main result of our work; the fate of the chiral anomaly in the Hofstadter regime.
Our starting point is Eq.~\eqref{eq:ham_mom} at $\Phi=0$ with chemical potential chosen to be at the Weyl nodes ($\mcl{E}_F=0$).
Upon applying a finite flux ($\Phi/\Phi_{0} \lesssim 1/4$)  the spectrum first breaks into Landau levels  (cf. Fig.~\ref{fig:ca_slopes}a) that disperse with $k_{z}$.
Due to the chiral anomaly, if an additional electric field is applied at $t=0$ satisfying $\mathbf{E}\parallel \hat{z}$, we expect the occupancies to shift along $k_z$ 
turning left-handed into right-handed fermions via the bottom of the band.
To characterize the chiral anomaly we define the chiral charge density
\eq{
\mcl{N}_5 = \left(\frac{1}{L^2}\right)\sum_{k_y=0}^{2\pi/a}\left[\sum_{k_z=0}^{\pi/a}n_{k_y,k_z} -\sum_{k_z=\pi/a}^{2\pi/a}n_{k_y,k_z}\right],
\label{eq:Nq}
}
where $n_{k_y,k_z}$ is the total number of filled fermionic states with momentum $k_y,k_z$.
This quantity monitors the amount of charge pumped from one half of the Brillouin zone to another; its rate of increase is proportional
to the applied electric field.
The definition~\eqref{eq:Nq} implies that only the states that cross the Fermi level can contribute to the pumping of chiral charge.
For $\Phi/\Phi_{0} \lesssim 1/4$ we find that $\mcl{N}_5$ grows linearly 
with time (Fig.~\ref{fig:ca_slopes}b).
In addition, the rate of growth $d\mcl{N}_5/Edt$ is linear as a function of $\Phi$  (Fig.~\ref{fig:ca_slopes}c). 
So far, both of these results are consistent with the conventional chiral anomaly, $d\mcl{N}_5/dt \propto \mathbf{E}\cdot\mathbf{B}$.

As the flux increased ($\Phi/\Phi_{0} > 1/4$) the linear behavior of $d\mcl{N}_5/Edt$ with $\Phi$ breaks down.
As shown in Fig.~\ref{fig:ca_slopes}c, several linear regimes where $d\mcl{N}_5/dt \propto \Phi$ appear, with unequal slopes.
Each linear regime is centered around commensurate fluxes $\Phi/\Phi_{0}=p/q$, with slope quantized to $q$.
This is a direct consequence of the emergence of $q$ pairs of Weyl nodes, leading to $q$ copies of the Landau levels crossing the Fermi energy.
Hence, as the flux is ramped, the Landau level degeneracy grows as $q\Delta \Phi$, leading to chiral charge production 
$\dot{\mcl{N}}_5 \propto E q \Delta \Phi$.
The full behaviour is thus composed of jumps between the linear regimes in Fig.~\ref{fig:ca_slopes} around commensurate values of the flux. 
In the thermodynamic limit the self similar fractal structure of the butterfly implies that these linear regimes should
themselves form a fractal of integer valued slopes.

In order to establish a more physical understanding of the fractal nature of the anomaly we 
now connect it to the Chern number. 
Recall first that the rate of  chiral charge pumping $\dot{\mcl{N}}_Q/E$ counts the number of chiral channels at the Fermi level. 
Second, we emphasize that the Weyl butterfly has, for fixed $k_z$, a series of gaps at $\mcl{E}_F=0$ (cf. Fig.~\ref{fig:ca_slopes}d-e),
each characterized by its Chern number $\mcl{C}_{k_{z}}$.
The Chern number determines how density is modified 
when applying a magnetic field through the Streda formula~\cite{Streda82}
\begin{equation}
{d \rho^{k_z}_\mathrm{2D}}/{dB_z} = {\mcl{C}_{k_z}}\vert_{\mcl{E}_F=0}/{\Phi_0}~.
\label{eq:weyl_streda}
\end{equation}
Consider adding one flux quantum to the system $\Phi=\Phi_0/L^2$. 
For $k_z=0$, $\mcl{C}=0$ at $E_F=0$, so the density is unaffected. 
For $k_z=\pi/a$, $\mcl{C}=1$,  so to increase $\rho_\mathrm{2D}$ as in Eq.~\eqref{eq:weyl_streda}, 
one conduction level must move to the valence band. 
The difference must be accommodated in between these momenta, leading to one extra chiral channel. 
%
\begin{figure}
 \includegraphics[width=\columnwidth]{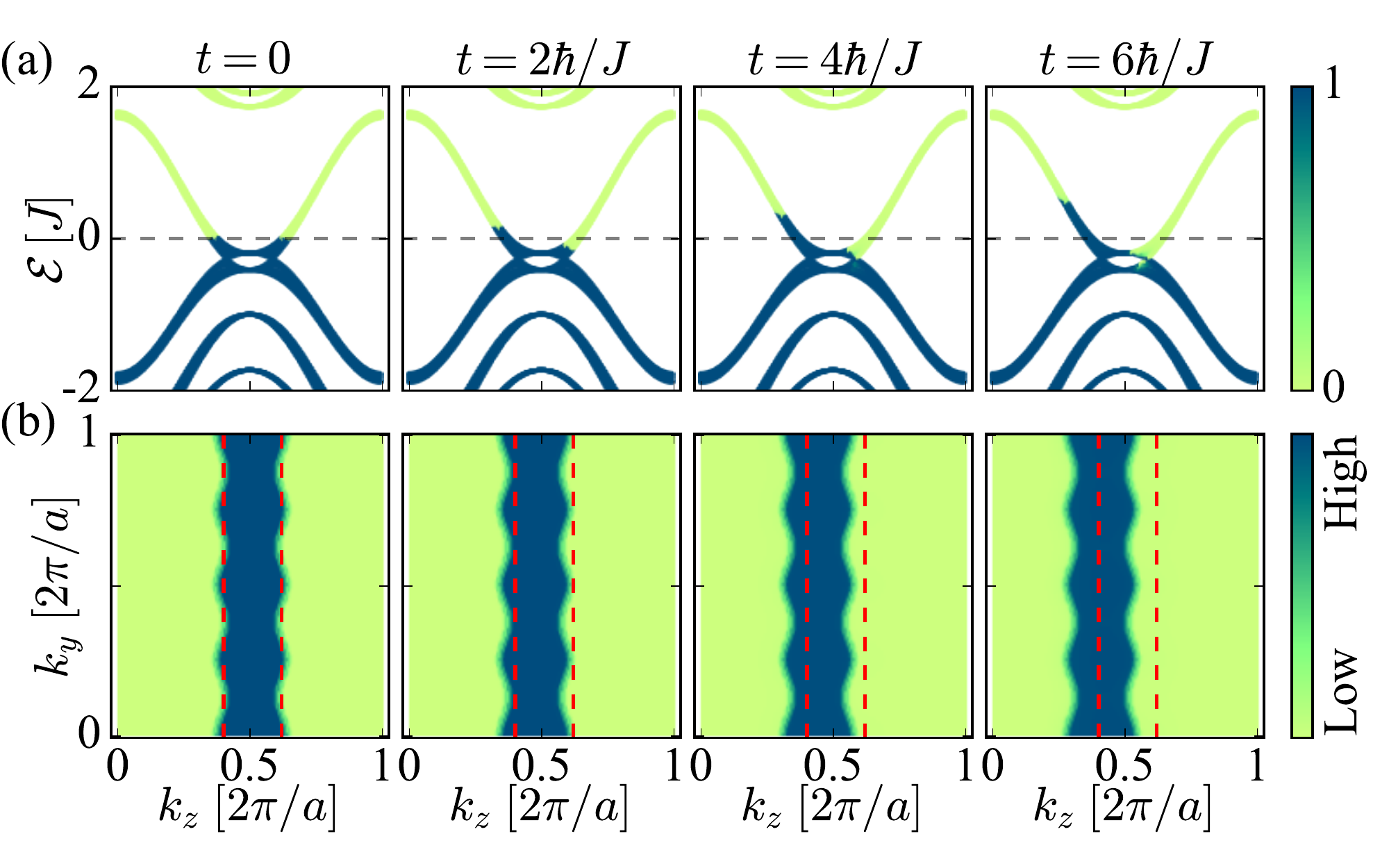}
 \caption{{\bf (a)} Evolution of occupancies in the bands after an electric field $E=0.1$ is turned on,
showing the development of the chiral anomaly for $\Phi{=}\Phi_0/4$. 
The dashed horizontal line represents the Fermi level $\mcl{E}_{F}$. 
{\bf (b)} Time-of-flight occupancy profiles in the $(k_y,k_z)$ plane in arbitrary units.
The vertical dashed lines serve as a guide to the eye for the initial profile at $t=0$.
Simulation parameters are the same as Fig.~\ref{fig:ca_slopes}.
}
 \label{fig:ca_tof}
\end{figure}
%
%
%
Since, for our inversion-symmetric Weyl semimetal, the Weyl points always appear in $\pm k_z$ pairs
with opposite chirality, it suffices to consider $k_{z}=0,\pi/a$. 
This predicts that the chiral anomaly generalizes to
\begin{equation}
\label{eq:Cdiff}
(1/E){d\dot{\mcl{N}}_5}/{ d\Phi} = \mcl{C}_{k_z=\pi/a}-\mcl{C}_{k_z=0} ~,
\end{equation}
which is confirmed in Fig.~\ref{fig:ca_slopes}c-e. Furthermore, since the butterfly at $k_z=\pi/a$ consists of a
fractal set of gapped Chern insulators, we see that the anomaly will become a fractal set of linear anomalies with 
quantized slopes in the thermodynamic limit.
Eq.~\eqref{eq:Cdiff} succinctly summarizes the main findings of this letter.
It highlights the topological connection between different $k_{z}$ sectors which determine the quantized slopes of the 
chiral anomaly, a result only possible in three-dimensions.

We close by addressing the experimental prospects to probe the chiral anomaly. Lack of 
reservoirs and relaxation make transport measurements difficult, but this also helps distinguish the chiral anomaly in
cold atoms from other competing effects.
In practice, the most direct probe is time-of-flight, which directly maps out the momentum-space occupancies.~\cite{BDZ2008}
Fig.~\ref{fig:ca_tof}a and Fig.~\ref{fig:ca_tof}b shows the calculated occupancies and time-of-flight images for $\Phi/\Phi_{0}=1/4$ upon applying $\mathbf{E} \parallel \mathbf{B}$
at $t=0$.
The pumping rate $\dot{\mcl{N}}_5$ can be monitored to probe the chiral anomaly and experimentally access the observables in Fig.~\ref{fig:ca_slopes}.

Our analysis extends to models without inversion symmetry, which may have multiple pairs of Weyl nodes.
In particular, time-of-flight measurements could track each pair of Weyl 
nodes independently to measure the chiral pumping.
Our results can thus be experimentally tested using existing technology in realistic models such as that in~Ref.~\cite{DKLKSB2015}, which already incorporates the
high magnetic field necessary for the Weyl butterfly,  or the three dimensional variant~\cite{WL16} of the model proposed in Ref.~\cite{LLN14}.
Finally, we expect these effects to be robust to weak interactions~\cite{GC11,IN12} and disorder~\cite{GPS15,KGM15,CSJ15}
leaving the effect of strong perturbations for future research.
\paragraph{Conclusion -- } 
We have shown that the chiral anomaly generalizes to a quantized fractal in the high-magnetic-field limit, 
connecting the longitudinal chiral anomaly response to the transverse Hall response
characterized by the Chern number. 
Our results hold for any model of Weyl semimetal with inversion symmetry.
The evolution of the spectral butterfly in the third momentum direction determines the universal three-dimensional physics of the chiral anomaly 
which is summarized by Eq.~\eqref{eq:Cdiff}.
This particular interplay between two-dimensional planes and the emergence of Weyl nodes for all $q$ distinguishes the Weyl butterfly
from two-dimensional~\cite{H76}, and three-dimensional variants of the Hofstadter problem~\cite{KHW92,H92,KAK01,KAO02,GK02,KA03,KA04,BDG04}
and opens the possibility of exploring generic features that relate different models.
%

%
\begin{acknowledgments}
\paragraph{Acknowledgements -- }
AGG acknowledges financial support from the European Commission under the
Marie Curie Programme. 
We thank C. Kennedy for useful insights on the experimental feasibility of this
proposal. 
MK and JEM were supported by Laboratory Directed Research and Development (LDRD) funding
from Berkeley Lab, provided by the Director, Office of Science, of the U.S. Department of Energy under Contract
No. DEAC02-05CH11231, and JEM acknowledges a Simons Investigator grant.
\end{acknowledgments}

\bibliography{WSM_WP}

\section{Supplementary material}

\subsection{Wavepackets with finite spread}

In this section, we illustrate how tuning the spread of wavepackets without changing the average momentum 
can be used to reconstruct the Chern number profile.
In particular, by tuning parameters such as temperature, density, trapping profile, and lattice
depth \cite{AEMWC95,DMAvDDKK95,BSTH95,GMEHB2002,GRJ2003,RGJ2004,JBAHRCHDG2003,ZSSRGHZ2003,BDZ2008}, cold atom
experiments routinely realize a wide variety of momentum-space distributions. Furthermore, the
distribution can not only be tuned via experimentally-accessible parameters, but measured with high 
accuracy in time-of-flight. As discussed in
the main text, this tunability of the momentum profile in the $z$-direction can be directly used 
to measure the Chern number profile as a function of $k_z$. In this appendix, we will show a particular
example of such a fitting procedure.

To model a wavepacket with finite momentum spread, we pick the simplest case of a Gaussian profile
centered around $k_z=0$ with width $\sigma$
\eq{
 W(k_z) = \frac{1}{\sqrt{2\pi}\sigma}\frac{e^{-k_z^2/2\sigma^2}}{\mathrm{Erf}(\pi/\sqrt{2}\sigma)},
\label{eq:Wkz}
 }
where the error function (Erf) results from normalizing the Gaussian distribution over the compact Brillouin zone.
The Hall drift in the $(x,y)$ plane 
would then yield a non-quantized effective Chern number given by the actual Chern number profile $\mathcal{C}_{k_z}$
averaged over the Brillouin zone and weighted by $W(k_z)$:
\eq{
\mcl{C}_{\mathrm{eff}} =\sum_{k_z=-\pi}^\pi W(k_z)~\mathcal{C}_{k_z}.
\label{eq:ceff}
}
Consider the case as in our model of the Weyl semimetal
where (possibly degenerate) Weyl nodes occur at $k_z=\pm K_0/2$.
The Chern number profile follows $\mcl{C}_{k_z} = \mcl{C}_{k_z=0}$ for $\vert k_z\vert <K_0/2$, and  $\mcl{C}_{k_z} = \mcl{C}_{k_z=\pi}$ otherwise (cf. Fig.~2 of the main text).
Hence, the problem has now been reduced to inferring three quantities, namely $K_0$, $\mcl{C}_{k_z=0}$ and
$\mcl{C}_{k_z=\pi}$, from measurements of $\mcl{C}_\mrm{eff}$ for various values of $\sigma$.

Taking a wavepacket that is well-localized near $k_z=0$ such as a Bose-Einstein condensate \cite{AEMWC95,DMAvDDKK95,BSTH95}
corresponds to $\sigma \approx 0$. 
Inspection of Eqs.~\eqref{eq:Wkz} and \eqref{eq:ceff}, trivially reveals that $\mcl{C}_\mrm{eff}(\sigma=0)=\mcl{C}_{k_z=0}$. Hence, the Hall drift with a wavepacket localized in $k_z$ would yield $\mcl{C}_{k_z=0}$, one of the three quantities of interest.

The other extreme limit, a ``high-temperature'' wavepacket completely delocalized in $k_z$ (while remaining
in the lowest band) has uniform distribution $W(k_z)=1/2\pi$. In this limit, $\mcl{C}_\mrm{eff}$ would saturate to 
\eq{
\mcl{C}_{\mathrm{sat}}=\mcl{C}_{k_{z}=\pi}-(\mcl{C}_{k_{z}=\pi}-\mcl{C}_{k_{z}=0})K_{0}/2\pi,
\label{eq:csat}
}
as mentioned in the main text.
$\mcl{C}_{\mathrm{sat}}$ relates the remaining two quantities of interest, $K_0$ and $\mcl{C}_{k_z=\pi}$.

\begin{figure}
\includegraphics[width=\columnwidth]{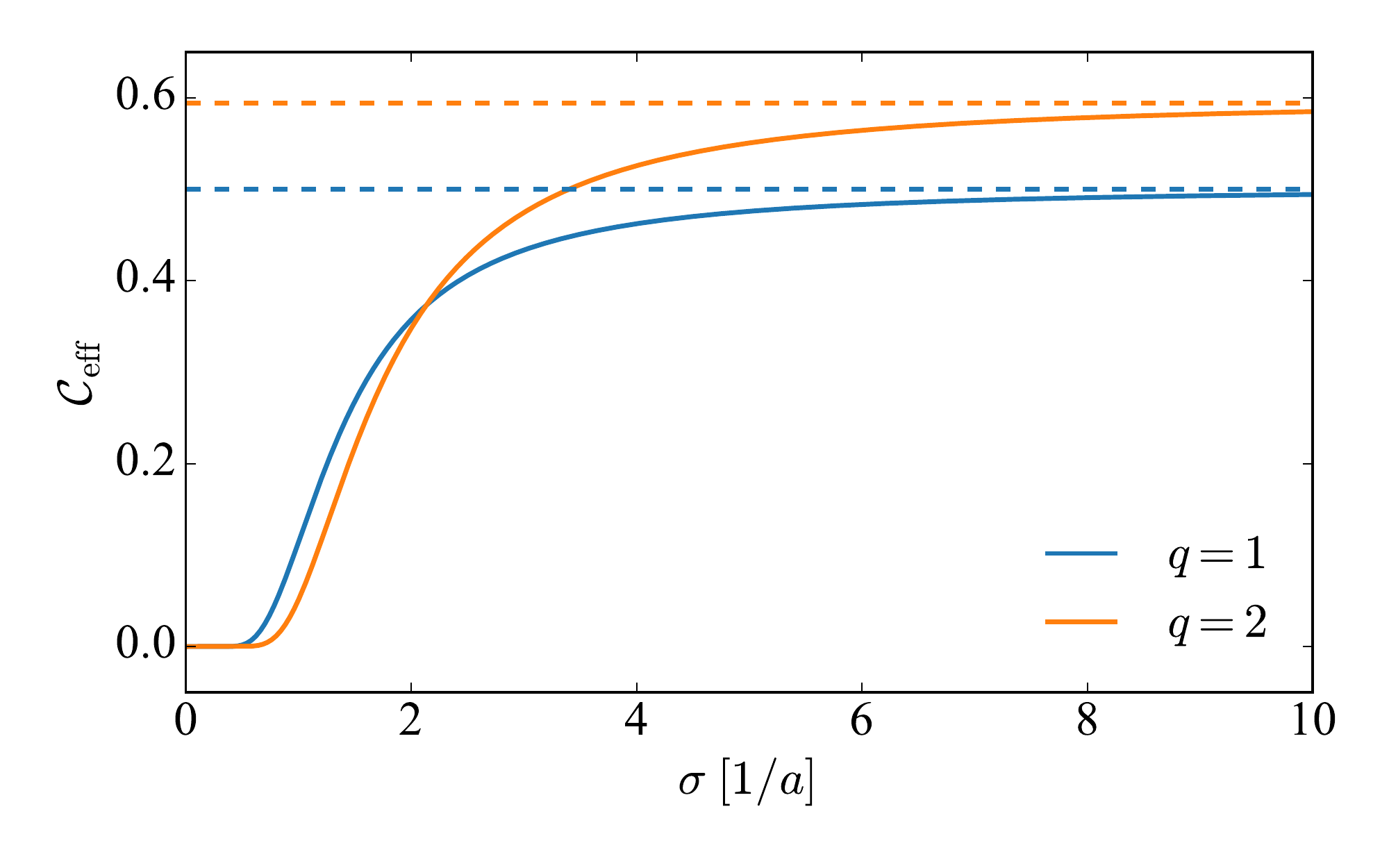}
\caption{$\mcl{C}_\mrm{eff}$ as a function of $\sigma$ for the two cases corresponding to Fig.~2 of the main text. The dashed lines correspond to the $\mcl{C}_{\mrm{sat}}$ values calculated from Eq.~\eqref{eq:csat}.}
\label{fig:fw}
\end{figure}

The behavior of $\mcl{C}_\mrm{eff}$ as a function of the width of the wavepacket in between 
these two limits is shown in Fig.~\ref{fig:fw}. 
For small $\sigma$ there is flat plateau at $\mcl{C}_{k_z=0}$
as the wavepacket has all its weight in the $\mcl{C}_{k_z}=\mcl{C}_{k_z=0}$ region of the Brillouin zone for $\sigma \ll K_0$.
Beyond this threshold, the tails of Gaussian \eqref{eq:Wkz} pick up contributions from the $\mcl{C}_{k_z}=\mcl{C}_{k_z=\pi}$ region and $\mcl{C}_\mrm{eff}$ deviates from the plateau.
Therefore, the length of this plateau is directly related to the separation of the Weyl nodes, $K_0$.
To determine this quantity it is useful to define empirically 
\eq{
\vert\mcl{C}_\mrm{eff}(\sigma_\Lambda) -\mcl{C}_{k_z=0}\vert = \Lambda,
\label{eq:cutoff}
}
where $\Lambda\ll \vert\mcl{C}_{k_z=0}-\mcl{C}_{k_z=\pi}\vert$, and $\sigma_\Lambda$ is the empirical length of the plateau.
Using the asymptotic properties of the error function, Eq.~\eqref{eq:cutoff} can be reduced to 
\eq{
\frac{2\sqrt{2}\sigma_\Lambda}{K_0}e^{-{K_0^2}/{8\sigma_\Lambda^2}} =  \Lambda \vert\mcl{C}_{k_z=\pi} -\mcl{C}_{k_z=0}\vert,
}
which can further simplified using Eq.~\eqref{eq:csat} to
\eq{
\frac{2\sqrt{2}\sigma_\Lambda}{K_0}\left(1-\frac{K_0}{2\pi}\right)e^{-{K_0^2}/{8\sigma_\Lambda^2}} =  \Lambda \vert\mcl{C}_{\mrm{sat}} -\mcl{C}_{k_z=0}\vert.
\label{eq:trans}
}
Note that $\mcl{C}_\mrm{sat}$ and $\mcl{C}_{k_z=0}$ are experimentally accessible quantities and $\sigma_\Lambda$ is an empirically chosen quantity. 
Hence, solving the transcendental equation \eqref{eq:trans} the value of $K_0$ can be obtained and used in Eq.~\eqref{eq:csat} to obtain $\mcl{C}_{k_z=\pi}$.
One may also perform a simple three parameter fit given the measured profile $W(k_z)$ and
obtain these parameters without any further analytical insight.

We have therefore shown that from experimentally-realistic procedures it is indeed possible 
to reconstruct the Chern number profile of a Weyl semimetal. 
We note that in the absence of interactions
these topological properties of the band structure are in principle equally accessible via boson
or fermions. Indeed, condensation has recently been measured in a two-dimensional 
Hofstadter model that can be thought of as a precursor to the Weyl semimetal \cite{KBCK2015}. Combined with 
wave packet measurements of Hall drift and Chern number that have been demonstrated in similar
two-dimensional models \cite{ALS14,JMD14}, these ideas for probing the topological physics of the Weyl semimetal
should be well within the reach of current technology.
\subsection{Weyl nodes at rational fluxes}
The fractal nature of the Weyl butterfly suggests that Weyl nodes may appear not just at some 
values of rational flux $\Phi=\Phi_0 p/q$, but for all rational fluxes. 
Our numerics confirm the emergence of $q$ pairs of Weyl nodes at all rational fluxes that we have checked.
In this section we provide a plausibility argument for this claim, though an analytic proof currently eludes us.

We start by noting that the effective 2D Hamiltonian obtained by considering the Weyl semimetal Hamiltonian (Eq. (1) in the main text) at a fixed $k_z$ slice can be decomposed into three kinds of terms,
the intra-sublattice hoppings ($J_1$), the inter-sublattice hoppings ($J_2$), and a staggered chemical potential $M_\mrm{eff}=M-J_1\cos k_z$.
For $J_2{=}0$, the Hamiltonian represents two decoupled square lattices; the spectra of each of these copies are shifted in energy by $\pm M_\mrm{eff}$. 

It was shown by Wen and Zee~\cite{WZ1989} that the single-particle spectrum of fermions hopping on a square lattice with a magnetic flux $\Phi$ per plaquette has at least $q$ isolated nodes at zero energy for $q$ even. 
This implies that the spectrum of Hamiltonian Eq. (1) at a fixed $k_z$ and $J_2=0$ has $2q$ nodes ($q$ nodes per sublattice) with $q$ located at $E=M_\mrm{eff}$ and $q$ at $E=-M_\mrm{eff}$.
However, a $J_2\ne0$ induces a mixing between the spectra of the two sublattices making the $q$ pairs of nodes gapped.
Such inter-sublattice hopping is analogous to a Haldane-like topological gap~\cite{H88} effectively breaking time-reversal symmetry.
As long as the two competing gaps, due to $M_\mrm{eff}$ and $J_2$,  are 
of the same order, the spectrum can be fine-tuned by varying $k_z$ to find gapless nodes in the spectrum.
The periodicity of the Brillouin zone ensures that there are $q$ of them 
and since the spectrum depends on $k_z$ via $\cos(k_z)$, 
the spectra at $\pm k_z$ are identical thus resulting in $q$-pairs of nodes.

\end{document}